# THE UTILIZATION OF SOCIAL NETWORKING AS PROMOTION MEDIA
# (CASE STUDY : HANDICRAFT BUSINESS IN PALEMBANG

**Dedi Rianto Rahadi[1], Leon Andretti Abdillah[2]**

[1]Postgraduate Program, Bina Darma University
Jl. Ahmad Yani No.12, Palembang, 30264
HP: +62 816 3288810
E-mail : dedi1968@yahoo.com[1]

[2]Computer Science Faculty, Information Systems Study Program, Bina Darma University
Jl. Ahmad Yani No.12, Palembang, 30264
HP: +62 813 6830 1000
E-mail : leon.abdillah@yahoo.com[2]

*Abstracts*

*Nowadays social media (Twitter, Facebook, etc.), not only simply as communication media, but also for promotion. Social networking media offers many business benefits for companies and organizations. Research purposes is to determine the model of social network media utilization as a promotional media for handicraft business in Palembang city. Qualitative and quantitative research design are used to know how handicraft business in Palembang city utilizing social media networking as a promotional media. The research results show 35% craft businesses already utilizing social media as a promotional media. The social media used are blog development 15%, facebook 46%, and twitter etc. are 39%. The reason they use social media such as, 1) minimal cost, 2) easily recognizable, 3) global distribution areas. Social media emphasis on direct engagement with customers better. So that the marketing method could be more personal through direct communication with customers.*

*Abstrak*

*Saat ini media sosial (Twitter, Facebook, etc.), tidak hanya sebagai media komunikasi, tapi juga untuk promosi. Media jejaring sosial menawarkan banyak keuntungan bisnis bagi perusahaan dan organisasi. Tujuan penelitian adalah untuk menentukan model pemanfaatan media jejaring sosial sebagai media promosi untuk usaha kerajinan di kota Palembang.desain penelitian kualitatif dan kuantitatif digunakan untuk menegetahui bisnis kerajinan tangan di Plaembang memanfaatkan jejaring sosial media sebagai media promosi. Hasil penelitian menunjukkan 35% bisnis kerajinan telah memanfaatkan media sosial sebagai media promosi. Media sosial yang digunakan adalah pengembangan blog 15%, facebook 46%, dan twitter dll 39%. Alasa mereka menggunakan media sosial adalah: 1) ongkos yang minimal, 2) mudah dikenali, 3) wilayah distribusi yang global. Media sosial menekankan pada keterlibatan langsung dengan pelanggan lebih baik. Sehingga metode pemasaran dapat lebih personal melalui komunikasi langsung dengan pelanggan.*

*Kata kunci: jejaring sosial media, pemasaran/promosi, hadiah bisnis*

## 1. INTRODUCTION

Information and communications technology (ICT) has developed very rapidly so as to provide a space for its users seamless access to information. Social networking media, is a media that is widely used to access the information. In the business world, the birth of social networking media can provide a huge benefit. There are so many businesses that has developed into a very good way just by using the right social media as a means of promotion and marketing.

Creating the perception of consumers about a product or service takes appropriate media. The use of appropriate media in order to make the products or services easily to be remembered and fostered positive perceptions for promoted products and services. Many campaign media can be used ranging from the common to the unique or new, for example, using the media or social networking. Social media or social nerworking sites such us

Facebook, witer, Blogs, etc. Could be used asa promotions media. These media are quite effective in obtaining new potential customers. Some social media has tools like social networking sites and instant messaging (Correa et al., 2010).

One of the social networking media is Facebook, which according to statistics reported by CheckFacebook.com, the number of Facebook users in Indonesia has entered big five, precisely number fourth, after U.S., Brazil, and India (Table 1).

Table 1. Facebook's users

| Rank | Country | Number |
| --- | --- | --- |
| 1. | United States | 159 436 740 |
| 2. | Brazil | 71 864 840 |
| 3. | India | 63 793 540 |
| 4. | Indonesia | 47 971 420 |
| 5. | Mexico | 42 944 260 |
| 6. | Turkey | 33 754 180 |
| 7. | United Kingdom | 31 130 240 |
| 8. | Philippines | 30 284 800 |
| 9. | France | 25 392 180 |
| 10. | Germany | 24 970 100 |

Source: CheckFacebook.com

SocialBakers also reported that for Indonesia, the demographics data show that he users disribution based on age are dominantly more then 50% of 18-24 years and 25-34 years (SocialBakers, 2013). There are 59% male users and 41% female users in Indonesia.

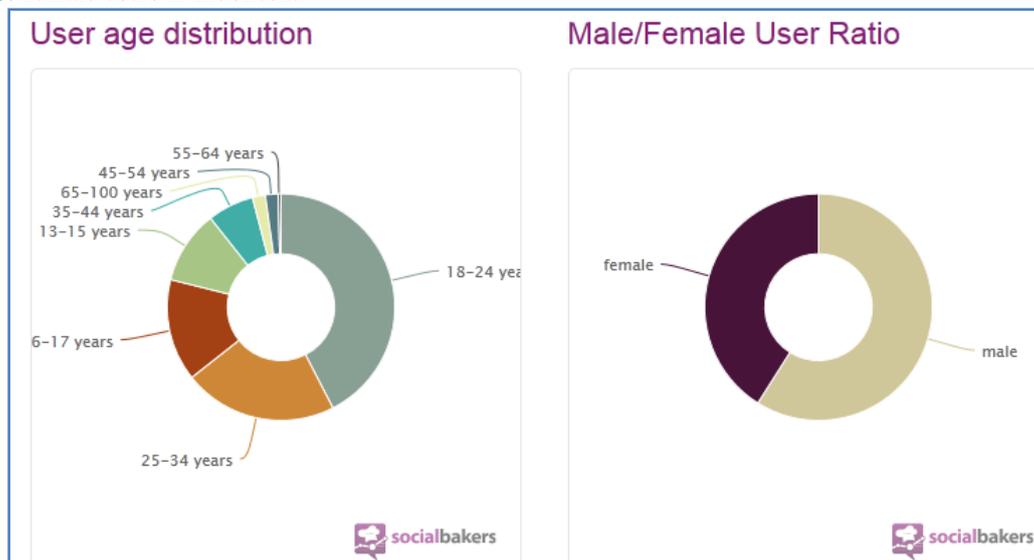

*Picture 1. Indonesian Facebook's Users Distribution*

## 2. RESEARCH METHODS

### 2.1 Data collection and Time of research

Qualitative and quantitative research methods are used to examine the use of social networking media as a promotional tool for craft businesses in the city of Palembang. Data in local reports reflect the brands' results as of the last day in the particular month, which they were generated. Engagement Rates and other metrics take into consideration the interactions that users created between the 1st and 31st of March 2013. If users like, comment or share a post beyond these dates, the interactions are not reflected in the metrics. That means that fan numbers as well as metrics from the local reports may slightly differ to the numbers users can currently see in Socialbakers Analytics.

43It turns out social networking or social media is likely to drive the growth of the Indonesian economy. Indonesian society today is one of the biggest consumer in the world of social media (Table 1). The amount of social networking users open new loopholes media marketing. These media do not require high IT expertise, because these media are basically only the items of goods sold through the company's image or the insertion of links to a variety of applications.

McQuail claimed as new media, social networking has several advantages (McQuail, 2010), such as: 1) *Interactivity*, 2) *Social presence (sociability),* 3) *Media richness*, 4) *Autonomy,* 5) *Playfulness,* and 6) *Personalization*.

First, Interactivity, the interactive nature of capabilities similar to the interactive capabilities interpersonal communication, the social networking company can Improve visibility / presence corporation By utilizing social networking company means more visibility / corporate presence. Can we assume the company opened a new show room.

Second, Social Presence (sociability), which is a great role to build a sense of personal contact with other participants of communication, here in the presence of social networks can facilitate the customers to contact us when they might lose the card company's name or phone number. Customers certainly remember the name of our company, then do a search on the internet, looking for our company, find and call us back.

Third, media richness, which is a bridge if there is a difference frame of reference, reduce ambiguity, giving cues, as well as more sensitive and more personal then social networking will improve the quality of customer service. In social networking, the company can easily add the information related to the products or services of the company. And can add information about the things that might solve the problems faced by customers. This information can be read directly or as a guideline for the customer support section. Thus easily obtain information about the customer complaints and support section also facilitated in serving its customers. Therefore the presence of social networking has the opportunity to improve customer satisfaction. So that a satisfied customer will continue to use the products or services of our company, we recommend the company to others and increase company revenue.

Fourth, Autonomy, that gives high freedom to users to manage the content and use. Through the media, communication participants can be independent of the communication resources. With the enterprise social networking can provide detailed company profile information at the time of an enterprise in relation to the other new enterprises he recognized (eg, recently received a supply), will probably find out more about the company profile recently gave offerings.

Fifth, playfulness, as entertainment and enjoyment, with the social networking will facilitate communication with corporate clients, friends, prospective customers and so on. Today, the modern social networking has been equipped with a variety of communication features that allow us to speak with the manager of the company.

Sixth, Personalization, it emphasizes that the contents of the message in the communication and its use is personal and unique. That is in addition to professional impression that a company that has a social network, it is definitely going to get more attention from companies that do not have a social network. Customers and prospective customers will certainly look first to the company that already has a social network. And can also increase the credibility of the company, where the company has a social networking will seem more advanced and modern. Thus, the credibility of the company will increase.

From the observations and interviews conducted by the researcher, described the problems that can arise are: 1) first, how to manage social networking media in order to keep uptodate information displayed, 2) second, how to introduce and promote the social networking media to be easily identifiable consumer.

### 2.2 Research framework

The research framework for this research could be seen on picture 2. Managing the promotional products and/or services through social media networking need a professional manager that focus on the controlling of the products/services as the content of publication.

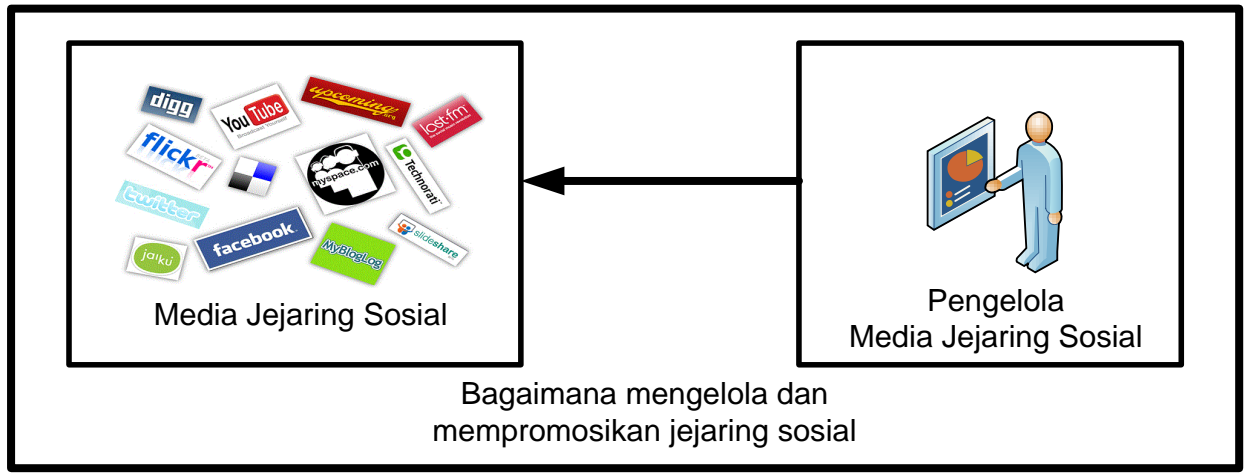

*Picture 2. Research framework*

### 2.3 Layout and specification

Qualitative and quantitative research design used to know and how to craft business in the city of Palembang utilizing social networking media as a promotional media. As informants and respondents is that carving craftsmen using social networking media in the city of Palembang. Examples of social networking media as respondent could be seen in picture 3.

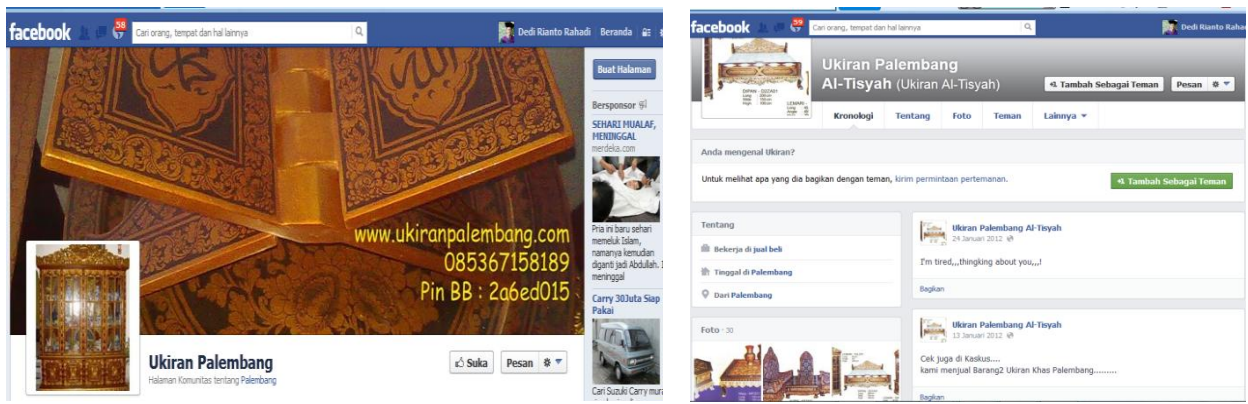

*Picture 3. Facebook as promotion media*

## 3. RESULTS AND DISCUSSIONS

### 3.1 Descriptive Statistics Analysis

The exploration investigation results undertaken on craftmen carving efforts using social networking media as a promotional media through interviews, surveys, questionaire dissemination, and be analyzed qualitatively and quantitatively show that:

First, 35% of craftmen businesses are utilizing social media industry as a media campaign. Social media used is 15% blog, 46% facebook, twitter and other media and others 39%. The reason they use social media such as, a) used relative costs are minimal so as not to burden the cost of production. This will result in a competitive selling price, b) could introduce branded products directly to customers, so easily recognizable, and c) range or distribution of a global sales. (Picture 4).

Second, the use of the Internet by using the social network has done since the last 2 to 3 years by 67% of respondents. This suggests that most of the respondents already use social networking as a promotion media for their products.

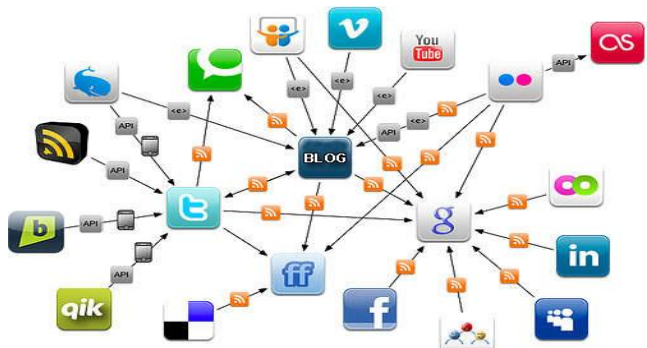

*Picture 4. Impact on Use of Social Media Networking*

Third, craftsman carving assume that social media especially facebook is very efficient, effective and inexpensive in the fulfillment of their needs. This is evidenced by 46 percent of respondents revealed the same thing. They feel that the particular needs through facebook shopping can save you a lot more time, effort and cost. Because all information required is very easy to access and there is a tendency to be more frugal shopping retailers.

Fourth, management of social networking is not done in a professional, craftsman utilizing only during free time so the impact of information displayed less attractive and not uptodate. Look less attractive, photos of little carvings displayed so that visitors to few social media networking.

Fifth, lack of transactions that occur due to lack of management focus in the use of social networking maximally. Impressed just be a formality that they have a social networking as a promotional media.

### 3.2 Discussions

The results showed the utilization of social networking is not optimal yet, this is due to the lack of professional management. As it is known that the increasing trend and the proliferation of the use of social media and the increasing number of users who can access the Internet, the more open opportunity society also business manager of creative economy and small and medium industries to reap the market through various social networking sites, thus can also spur economic growth in Indonesia.

Use the media that popular and recognize by many pepole to develop the image and increase the popularity of products. We also nee to maintain the quality of our brand in order to satisfy the customers. We also need to open the space for our customers to give any feedbacks or critiques for the quality of our products via popular social networking media.

Social Media with its power become the powerfull tools for every aspects, besides promotion and marketing, social media also able become self imaging media became a powerful tool in a variety of ways, in addition to the promotion of social media marketing distinction can also be a means of self-image, it is banyaak done by figures in the political world in gaining public sympathy. So formidable power of social media revolution in the world some countries also initiated and led by the movement and spread through social media.

One study by Chadwick Martin said that 33 percents of facebook's users are fan of brands and products, (Owyang et al., 2010), where 60 percents of those consumers will recommend the brands or procusts to their friends. This finding confirmed the previous study by Marketing Sherpa (2009) that highligheting the consumers in which highlights that consumers in meeting the need for product information or to refer to what has been done by other facebook users.

Another report by Harris in 2011 mention that consumers are also bringing their online experiences into their own social networks rather than engaging directly on company websites (offsite social commerce) (Harris and Dennis, 2011). Furthurmore they argued that social networking, particularly Facebook, is becoming ever more prevalent, particularly with young people.

To take advantage of social media to promote products or services that we produce, then there are some things we must note that the purpose of promotion that can be achieved, namely: 1) Review the products and services marketed by the model we wish continued, 2) Spread the good information and useful for many people so that people will always follow and read what we write in our social media include reading a variety of products and services that we promote, 3) Avoid using harsh language and hostile use of social media that we have, 4) Interact with interactive and friendly, 5) Create a blog or a website that reviews about what we are doing, including the advantages of the products and services we market, 6) Take advantage of social media to create what word of mouth marketing word of mouth, 7) Create a good image for the products we have to offer, 8) Processing the words correctly, and 9) Take advantage of email marketing as a promotional tool one by one. Utilizing Yahoo Messenger, Facebook and twitter to interact with customers. Even in Twitter if what we are promoting could enter the top trending topic then there will be more people who know our business is being run.

For self development, never stop to learn and join various seminar about the utilization of social media as promotion media. There are many tips and tricks that we could get from those seminar that will support us in using social media for the improvement of our business.

## 4. CONCLUSIONS AND SUGGESTIONS

### 4.1 Conclusions

This exploratory investigation results indicate that the use of social networking media has been done partly craftsman carving, but the feedback are not optimal yet. The absence of social networking media managers who focus on that job affects the quality of the content and information displayed. Craftsmen feel the benefits of social networking /facebook to fulfill all the needs including communication needs, promotional products, customer relationships, as well as other needs. Social networking has been very effective and efficient in meeting the needs of promotional products. In the sense that much more information can be obtained, more time-saving, labor-saving and cost-effective.

### 4.2 Suggestions

It takes the seriousness in managing social media networks in order to provide optimal results particularly in promoting the product. Join the business communities to introduce existing social networking so that more visitors will visit it.